\documentclass[aps,prl,twocolumn,amsmath,amssymb,superscriptaddress,nofootinbib]{revtex4-1}
\usepackage{url}

\usepackage{epsfig}
\usepackage{amsfonts}
\usepackage{graphicx}
\usepackage{epsfig}
\usepackage{eepic}
\usepackage{amsmath}
\usepackage{amssymb}
\usepackage{color}
\usepackage{bbm}
\usepackage{dcolumn}
\usepackage{bm}
\usepackage{ulem}
\usepackage{mathrsfs}
\usepackage{bbold}
\usepackage{datetime}

\usepackage{overpic} 
\usepackage{rotating}
\usepackage[usenames,dvipsnames]{xcolor}
\usepackage[colorlinks=true,citecolor=Magenta,linkcolor=Green,urlcolor=Blue]{hyperref}
\usepackage{lipsum} 
\usepackage{etoolbox} 
\usepackage[capitalize]{cleveref}

\def\bc{\begin{center}}

\def\ec{\end{center}}
\def\be{\begin{eqnarray}}
\def\ee{\end{eqnarray}}
\definecolor{dyellow}{rgb}{1.,0.8,.0}
\definecolor{myblue}{rgb}{.1,.1,.7}
\definecolor{dcyan}{rgb}{.0,.6,.6}
\definecolor{dmagenta}{rgb}{0.6,0.0,0.6}
\definecolor{brown}{rgb}{0.6,0.2,0.}
\definecolor{darkblue}{rgb}{.0,.0,0.5}
\definecolor{darkred}{rgb}{0.75,0.0,0.0}
\definecolor{orange}{rgb}{1.,.6,.0}
\definecolor{dorange}{rgb}{0.8,.4,.0}
\definecolor{darkgreen}{rgb}{0.0,0.6,0.0}
\definecolor{purple}{rgb}{.4,.0,.4}
\definecolor{lightgrey}{rgb}{0.7, 0.7, 0.7}
\definecolor{grey}{rgb}{0.4, 0.4, 0.4}

\usepackage[font=small,labelfont=bf,justification=raggedright]{caption}

\newcommand{\xdownarrow}[1]{%
  {\left\downarrow\vbox to #1{}\right.\kern-\nulldelimiterspace}
}
\newcommand{\xuparrow}[1]{%
  {\left\uparrow\vbox to #1{}\right.\kern-\nulldelimiterspace}
}

\begin{document}
\newsavebox{\lefttempbox}
\title{\bf \Large Holographic topological defects and local gauge symmetry: clusters of strongly coupled equal-sign vortices}

\author{Zhi-Hong Li}\email{lizhihong@buaa.edu.cn}
\affiliation{Center for Gravitational Physics, Department of Space Science, Beihang University, Beijing 100191, China}

\author{Chuan-Yin Xia}\email{xiachuanyin@163.com}
\affiliation{Department of Electronic Science and Applied Physics, School of Science, Kunming University of Science and Technology, Kunming 650500, China}
\affiliation{Center for Gravitation and Cosmology, School of Physics Science and Technology, Yangzhou University, Yangzhou 225009, China}

\author{Hua-Bi Zeng}\email{hbzeng@yzu.edu.cn}
\affiliation{Center for Gravitation and Cosmology, School of Physics Science and Technology, Yangzhou University, Yangzhou 225009, China}

\author{Hai-Qing Zhang}\email{hqzhang@buaa.edu.cn}
\affiliation{Center for Gravitational Physics, Department of Space Science, Beihang University, Beijing 100191, China}
\affiliation{International Research Institute for Multidisciplinary Science, Beihang University, Beijing 100191, China}

\begin{abstract}
{Gauge invariance plays an important role in forming topological defects. In this work, from the AdS/CFT correspondence, we realize the clusters of equal-sign vortices during the course of critical dynamics of a strongly coupled superconductor. This is the first time to achieve the equal-sign vortex clusters in strongly coupled systems.  The appearance of clusters of equal-sign vortices is a typical character of flux trapping mechanism, distinct from Kibble-Zurek mechanism which merely presents vortex-antivortex pair distributions resulting from global symmetry breaking.  Numerical results of spatial correlations and net fluxes of the equal-sign vortex clusters quantitatively support the positive correlations between vortices. The linear dependence between the vortex number and the amplitude of magnetic field at the `trapping' time demonstrates the flux trapping mechanism very well.
 }

\end{abstract}

\maketitle

\clearpage

Formation of topological defects due to global symmetry breaking in a phase transition is generically described by the Kibble-Zurek mechanism (KZM) \cite{Kibble:1976sj,Zurek:1985qw}.  It states that during a continuous phase transition, global symmetry breaking will occur inside some causally uncorrelated regions (freeze-out regions) because of the critical slowing down of the order parameter near the critical point. Topological defects may form with some probabilities between those adjacent regions \cite{Bowick:1992rz,delCampo:2018hpn}. Therefore, the number density of defects can be estimated from the critical dynamics of the theory. KZM has been tested in many numerical simulations and experiments, such as in superfluids \cite{Baeuerle:1996zz,Ruutu:1995qz}, liquid crystals \cite{Bowick:1992rz,Chuang:1991zz,Digal:1998ak} and quantum optics \cite{Guo} (for reviews, see \cite{Kibble:2007zz,delCampo:2013nla}).

While most of previous research focused on systems with global symmetry, it is useful to explore the systems with local gauge symmetry \cite{Hindmarsh:2000kd,Stephens:2001fv}. In this case, the defects formation are distinct from KZM since the local phase gradients can be removed by the gauge transformations. This would lead to new phenomenology, in which the underlying physics is dubbed ``flux trapping mechanism" (FTM) \cite{Kibble:2003wt}. Consequently, the resulting defects number will be proportional to the relevant magnetic fluxes at the `trapping' time. In the past two decades, FTM has been studied in superconducting films \cite{donaire2007,Kireley2003} and cosmology \cite{BlancoPillado:2007se}.

The key difference from KZM and FTM is the spatial distribution of the defects stemming from distinct correlations between them \cite{Hindmarsh:2000kd,Stephens:2001fv}. In KZM, random choices of order parameter phases in the freeze-out regions lead to negative correlations between defects, i.e, they are distributed in defect-antidefect pairs at short range.  However, in FTM the defects are positively correlated. In other words, those defects should be formed in clusters of equal sign. Numerical simulations of these clusters have been realized already in \cite{Stephens:2001fv,donaire2007,BlancoPillado:2007se} for weakly coupled systems.

However, the equal-sign vortex clusters have not been studied in strongly coupled field theory. We will investigate it by virtue of AdS/CFT correspondence. AdS/CFT correspondence, which is a ``first-principle'' route to solving strongly coupled physics, comes to rescue \cite{Maldacena:1997re,Zaanen:2015oix}. In this letter, we investigate the defects formation with local gauge symmetry breaking by utilizing the AdS/CFT technique.  We add a plane-wave magnetic field in the initial state, as in \cite{Rajantie:2001na}.  Quenching the system linearly through the critical point, order parameter vortices and the related quantized magnetic fluxes (fluxoids) are spontaneously generated. Since this is a type-II superconductor \cite{Zeng:2019yhi,hartnoll}, order parameter vortices are confined into the fluxoids. Clusters of equal-sign vortices turn out with positive (negative) vortices packing together in the regions of initial positive (negative) magnetic fields. The corresponding spatial correlation function has a positive maximum, indicating a positive correlation between vortices. Net flux of vortices inside a closed area quantitatively supports the conclusions of positive correlations above. Numerically, we find a linear dependence between the vortex number and the amplitude of the magnetic field at the `trapping' time, verifying the FTM very well. Previous work on holographic topological defects can be found in \cite{Zeng:2019yhi,Chesler:2014gya,Sonner:2014tca,Li:2019oyz,delCampo:2021rak,Li:2021iph}.

\section{Basic setup}
\label{background}
{\bf Background of gravity:}
 The gravity background is the AdS$_4$ black brane in Eddington-Finkelstein coordinates,
\begin{equation}
ds^2 = \frac{L^2}{z^2} (-f(z) dt^2 - 2dtdz + dx^2 + dy^2),
\end{equation}
where $f(z) = 1 - (z/z_h)^3$, with $\{L, z, z_h\}$ representing the AdS radius, AdS radial coordinate and the location of horizon respectively. The AdS infinite boundary is at $z = 0$ where the field theory lives.  Lagrangian of the model we adopt is the usual Abelian-Higgs model for holographic superconductors \cite{hartnoll},
\begin{equation}\label{density}
\mathcal{L} = -\frac{1}{4} F_{\mu \nu} F^{\mu \nu} - |D \Psi|^2 - m^2 |\Psi|^2.
\end{equation}
where $\Psi$ is the complex scalar field and $D=\nabla -iA$ is the covariant derivative with $A$ the U(1) gauge field (we have imposed the electric coupling constant $e\equiv1$). We work in the probe limit, then the equations of motion read,
\begin{eqnarray}\label{eomofwhole}
D_\mu D^\mu\Psi-m^2\Psi=0, \nabla_\mu F^{\mu\nu}=i\left(\Psi^* D^\nu\Psi-\Psi{(D^\nu\Psi)^*}\right). ~~~~
\end{eqnarray}
The ansatz we will take is $\Psi = \Psi(t,z,x,y), A_{t,x,y} = A_{t,x,y}(t,z,x,y)$ and $A_z = 0$.

\begin{figure}[t]
\centering
\includegraphics[trim=1cm 0.8cm 0.8cm 2.3cm, clip=true, scale=0.3, angle=90]{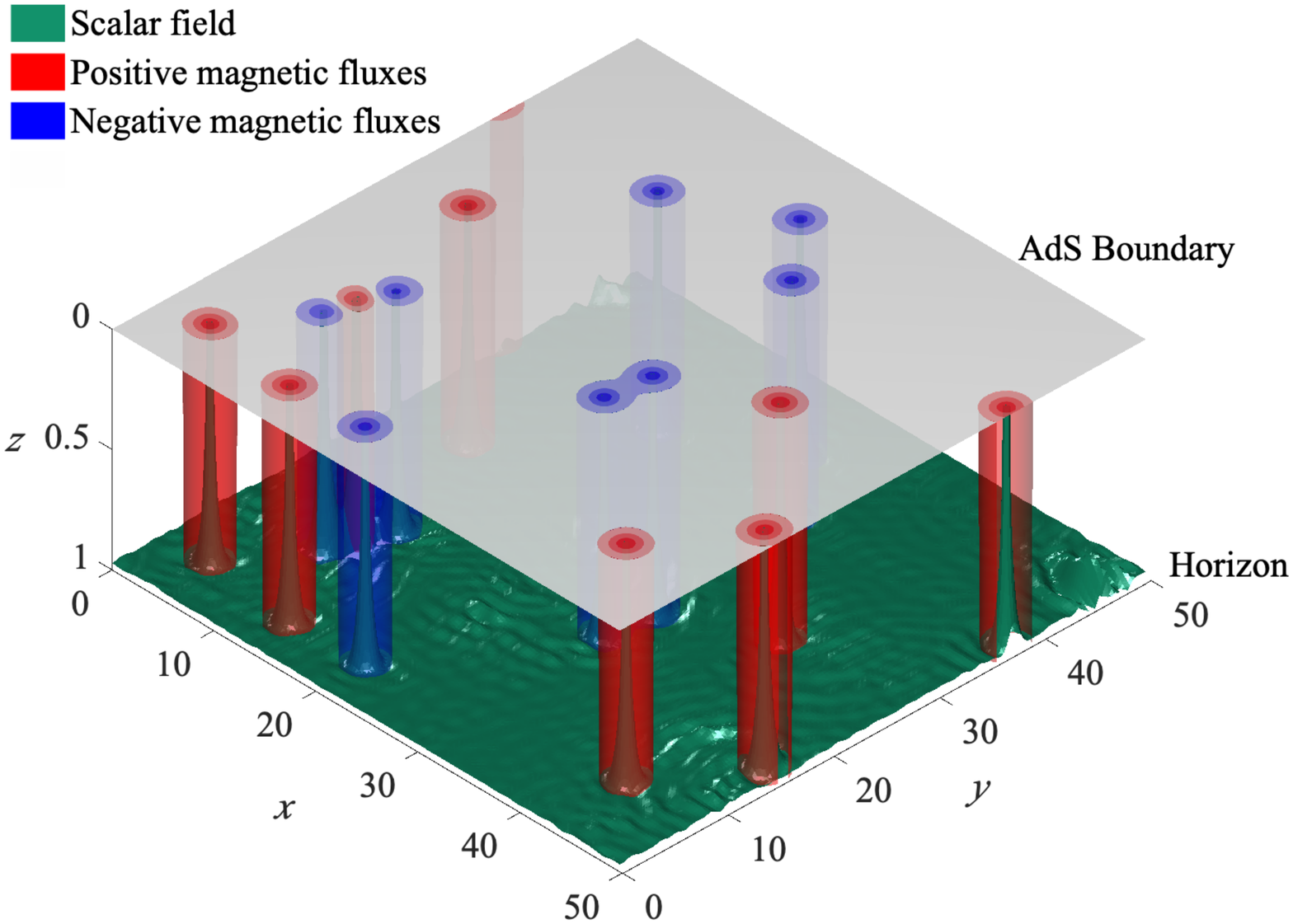}
\caption{Typical configurations of matter fields in the bulk at the final equilibrium state (the average order parameter arrives at a plateau) after quench. Green color is the isosurface of scalar field while red (blue) color represents positive (negative) magnetic fluxoids. Order parameter vortices are confined in the magnetic fluxoids, indicating a type-II superconductor system. }\label{3D} 
\end{figure}

{\bf Boundary conditions \& holographic renormalization:} The asymptotic behaviors of fields near $z\to0$ are $A_\mu\sim a_\mu+b_\mu z+\dots, \Psi=\frac zL\left(\Psi_0+\Psi_1 z+\dots\right)$. We have set the scalar field mass square as $m^2= -2/L^2$. In the numerics we have scaled $L=1$.
 From AdS/CFT correspondence, $a_t, a_i~ (i=x, y)$ and $\Psi_0$  are interpreted as the chemical potential, gauge field velocity and source of scalar operators on the boundary, respectively. Their conjugate variables can be evaluated by varying the renormalized on-shell action $S_{\rm ren}$ with respect to these source terms.  From holographic renormalization \cite{Skenderis:2002wp}, the counter term for the scalar field is $S_{\rm ct}=\int d^3x\sqrt{-h}\Psi^*\Psi$, where $h$ is the reduced metric on the $z\to0$ boundary. In order to have dynamical gauge fields in the boundary, we need to impose Neumann boundary conditions for the gauge fields as $z\to0$ \cite{witten,silva}. Therefore, the surface term $S_{\rm surf}=\int d^3x\sqrt{-h}n^\mu F_{\mu\nu}A^\nu$ for the gauge fields should also be added in order to have a well-defined variation, where $n^\mu$ is the normal vector perpendicular to the $z\to0$ boundary. Finally, we obtain the finite renormalized on-shell action $S_{\rm ren}$. Therefore, the expectation value of the order parameter $\langle O\rangle=\Psi_1$, can be obtained by varying $S_{\rm ren}$ with respect to $\Psi_0$. Expanding the $z$-component of the Maxwell equations near boundary, we get $\partial_tb_t+\partial_iJ^i=0$. This is exactly a conservation equation of the charge density and current on the boundary, since from the variation of $S_{\rm ren}$ one can easily obtain $b_t=-\rho$ with $\rho$ the charge density and $J^i=-b_i-(\partial_ia_t-\partial_ta_i)$ which is the $i$-direction current respectively.

On the $z\to0$ boundary, we set $\Psi_0=0$ in order to have spontaneous symmetry breaking, and this gives rise to a non-vanishing order parameter. The Neumann boundary conditions for the gauge fields are imposed from the above conservation equations. Therefore, dynamical gauge fields on the boundary can be evaluated and lead to the spontaneous formation of magnetic fluxoids. Moreover, we impose the periodic boundary conditions for all the fields along $(x, y)$-directions. At the horizon we set $A_t(z_h)=0$ and the regular finite boundary conditions for other fields.

{\bf Cool the system:} From dimension analysis, temperature of the black hole $T$ has mass dimension one, while the mass dimension of the charge density $\rho$ is two. Therefore, $T/\sqrt{\rho}$ is dimensionless. From holographic superconductor \cite{hartnoll}, decreasing the temperature is equivalent to increasing the charge density. Therefore, in order to linearly decrease the temperature as $T(t)/T_c=1-t/\tau_Q$ near the critical point conventionally \cite{Zurek:1985qw} ($\tau_Q$ is called the quench rate), one can indeed quench the charge density $\rho$ as
\begin{eqnarray}\label{quench}
\rho(t)=\rho_c\left(1-t/\tau_Q\right)^{-2}
\end{eqnarray}
where $\rho_c$ is the critical charge density for the static and homogeneous holographic superconducting system. A typical configuration of the holographic system is exhibited in Fig.\ref{3D}, which is obtained in the final equilibrium state after quench. \footnote{In our paper, the `equilibrium state' refers to the state when the average order parameter arrives at a plateau with vortices, rather than the thermal equilibrium state without any vortices.}

\begin{figure*}[t]
\centering
\includegraphics[trim=4.2cm 6.cm 5.2cm 0.7cm, clip=true, scale=0.5, angle=0]{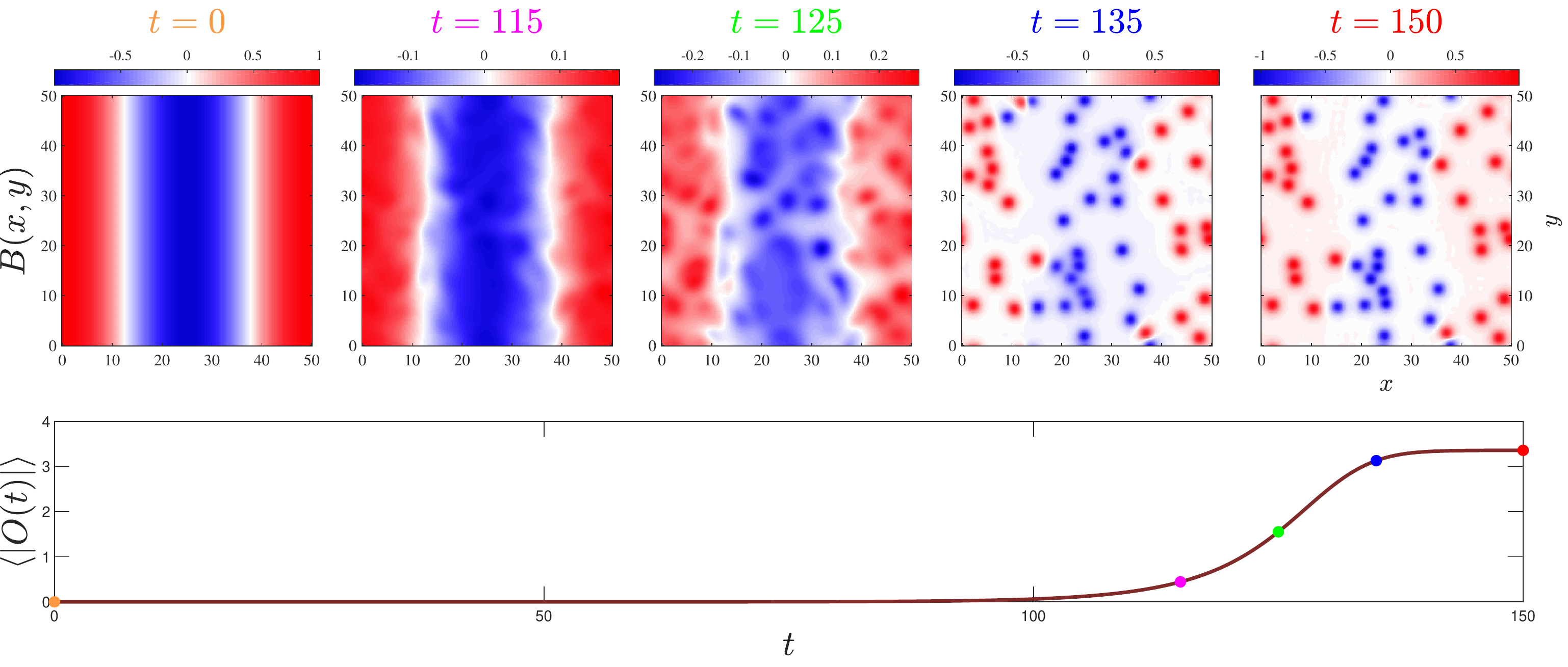}
\put(-458,180){(a)}
\put(-458,64){(b)}
\caption{{Time evolution of the magnetic field and the birth of topological defects.} (a) Density plots of the evolving magnetic field at five specific times ($t=0, 115, 125, 135, 150$) with $\tau_Q=20$ and $B_0=1$. One can see how the initial plane-wave magnetic field evolves to the clusters of equal-sign vortices.  The red (blue) localized points in the equilibrium state ($t=150$) represent positive (negative) magnetic fluxoids; (b) The growth of the average order parameter from initial time to final equilibrium state. Five colored points correspond to the five snapshots in panel (a), respectively. The order parameter scrambles during the period $t\approx [100, 135]$.  Related animations can be found in the movie \href{https://bhpan.buaa.edu.cn:443/link/95BA4D4BD5B130767C9012BFAC4C91DB}{M1.avi}. }\label{B}
\end{figure*}

{\bf Numerical schemes:}
The system evolves by using the 4th order Runge-Kutta method with time step $\Delta t=0.1$.\footnote{ We have also checked that smaller time steps will lead to similar numerical results.  For example performing more than one run with time step $\Delta t=0.05$ and $\Delta t=0.01$, the results are similar to $\Delta t=0.1$. Therefore, our numerical results with $\Delta t=0.1$ are reliable. In order to run faster, we set $\Delta=0.1$ which will not ruin the accuracy of the results.} In the radial direction $z$, we used the Chebyshev  pseudo-spectral method with 21 grid points. Since in the $(x, y)$-directions all the fields are periodic, we use the Fourier decomposition along $(x, y)$-directions with $201\times201$ grid points. Filtering of the high momentum modes are implemented following the ``$2/3$'s rule'' that the uppermost one third Fourier modes are removed \cite{Chesler:2013lia}.

\section{Results}

{\bf Formation of clusters of equal-sign vortices:}   Distinct from the settings in \cite{Hindmarsh:2000kd,Stephens:2001fv}, we adopt a simple and instructive form of magnetic field at the initial time, which may be operated with techniques of magnetic fields in experiments \cite{borisenko2020}. Specifically, we add a plane-wave magnetic field along $x$-direction at the initial time $t_i$ as \cite{Rajantie:2001na},
\begin{eqnarray}\label{mag}
B(x,y)\big|_{t=t_i}=B_{0}\cos\left(kx\right).
\end{eqnarray}
where $B_0$ is the initial amplitude of magnetic fields while $k$ is the wave number. Because of $B=\partial_{x}A_{y}-\partial_{y}A_{x}$, one can set the initial condition of $A_x$ and $A_y$ as $A_x(t=t_i)=0, A_y(t=t_i)=\frac{B_{0}}{k}\sin\left(kx\right)$. Obviously this magnetic field is perpendicular to the AdS boundary $z=0$. Without loss of generality, we choose $k=2\pi/l$ where $l$ is the length of each side of the $(x,y)$ boundary and we impose $l=50$. We already checked that other choices of $k$ will also obtain similar results. We quench the system from the initial temperature $T_i=T_c$ to the final temperature $T_f=0.8T_c$, and then maintain the system at $T_f$ until it arrives at the equilibrium state. \footnote{Previous work \cite{Stephens:2001fv, Huang} already showed that it was viable to start the quench near $T_c$ rather than much greater than $T_c$, since they found that the symmetry-breaking actually occurred after crossing the critical point. Based on their results in weakly coupled systems, we assume that they are also applicable in strongly coupled systems. }

Fig.\ref{B} shows the evolution of the magnetic field (panel (a)) and the average order parameter (panel (b)) from the starting of quench to the final equilibrium state with quench rate $\tau_Q=20$. Panel (a) shows five snapshots at times $t={0, 115, 125, 135, 150}$, corresponding to the five colored points in the panel (b), respectively.  At the initial time the shape of the magnetic field is a plane wave as Eq.\eqref{mag}. As quench initiates, the amplitude of magnetic field will exponentially decay until the order parameter is relatively large (see the \hyperref[Bdecay]{Appendix} for details).
When system enters the superconducting phase, magnetic fields will be forbidden by the Meissner effect. However, because of causality, the magnetic fields are unable to decay if the phase transition takes place in a finite time. Then, magnetic fields survive even after the transition, and the only way they can do is to generate quantized magnetic fluxoids.
At a later time $t=115$, the magnetic field is no longer in the plane-wave shape, while the order parameter gets bigger and is in the ramping stage from panel (b).  As the order parameter climbs to the middle stage of the ramp ($t=125$), numerous lumps occur in the magnetic field in panel (a). This is due to the Meissner effect in the superconductor. The lumps are actually the concentrate of magnetic fluxes where the cores of vortices will finally locate (see the \hyperref[location]{Appendix} for details). Meissner effect suppresses the magnetic field surrounding these lumps. As time goes by, when the order parameter just arrives at the equilibrium state ($t=135$), blue (red) islands of lumps finally form clusters of negative (positive) magnetic fluxoids. These snapshots intuitively show the FTM of how the clusters of equal-sign vortices form.

\begin{figure}[h]
\centering
\includegraphics[trim=3.cm 8cm 3.5cm 8cm, clip=true, scale=0.5]{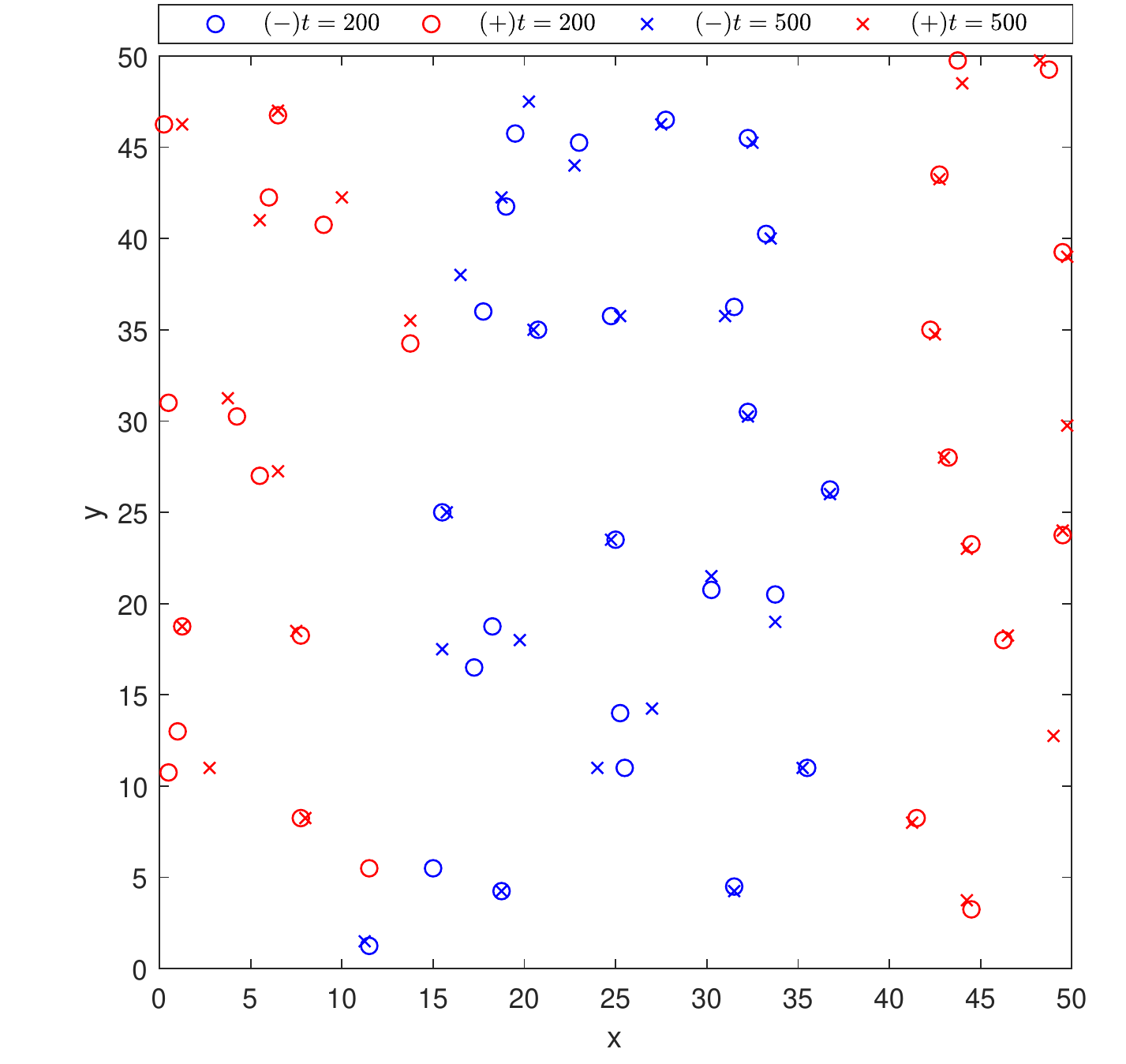}
\caption{ The positions of magnetic fluxoids at different times ($t=200$ and $t=500$) with $\tau_Q=20$ and $B_0=1$ from the attached movie \href{https://bhpan.buaa.edu.cn:443/link/D016990DD26F033F72A719CE02F9CD08}{M2.avi}. The circles indicate the positions of the magnetic fluxoids at $t=200$, and the crosses indicates the positions of the magnetic fluxoids at $t=500$. Red color represents positive magnetic fluxoids and blue color represents negative magnetic fluxoids.} \label{lo}
\end{figure}

From Fig.\ref{B}, we see that keeping the system in the equilibrium until $t=150$, most vortices hardly move except a pair of nearby negative and positive vortices annihilate at the position $(x\approx 12, y\approx49)$.  This phenomenon is reminiscent of the ``pinning effect'', which is a typical phenomenon in type-II superconductor if there exists the magnetic fluxes \cite{tinkham}. In order to illustrate the pinning effect clearly, we demonstrate the dynamical process of vortices in holographic superfluids and holographic superconductor in the movie \href{https://bhpan.buaa.edu.cn:443/link/D016990DD26F033F72A719CE02F9CD08}{M2.avi}. In this movie, the left column is about the holographic superfluids while the right column is for holographic superconductors. As times goes by, vortices in holographic superfluids will move closely and then all annihilate eventually. However, most vortices in holographic superconductors will almost stay in their original places, only very few vortices will move together and then annihilate.  In Fig.\ref{lo}, we plot the positions of magnetic fluxoids in holographic superconductors at different times ($t=200$ and $t=500$) excerpted from this movie \href{https://bhpan.buaa.edu.cn:443/link/D016990DD26F033F72A719CE02F9CD08}{M2.avi}. We can see clearly that most magnetic fluxoids will almost stay in their original places during the time scale that the vortices in holographic superfluids will all annihilate. Only a pair of vortices at locations $(x\approx11,y\approx5.5)$ and $(x\approx15,y\approx5.5)$ will move together and then annihilate. We leave the studies of the details of the pinning effect as a future work.

\begin{figure}[t]
\centering
\includegraphics[trim=3.2cm 9.5cm 4cm 9.5cm, clip=true, scale=0.5]{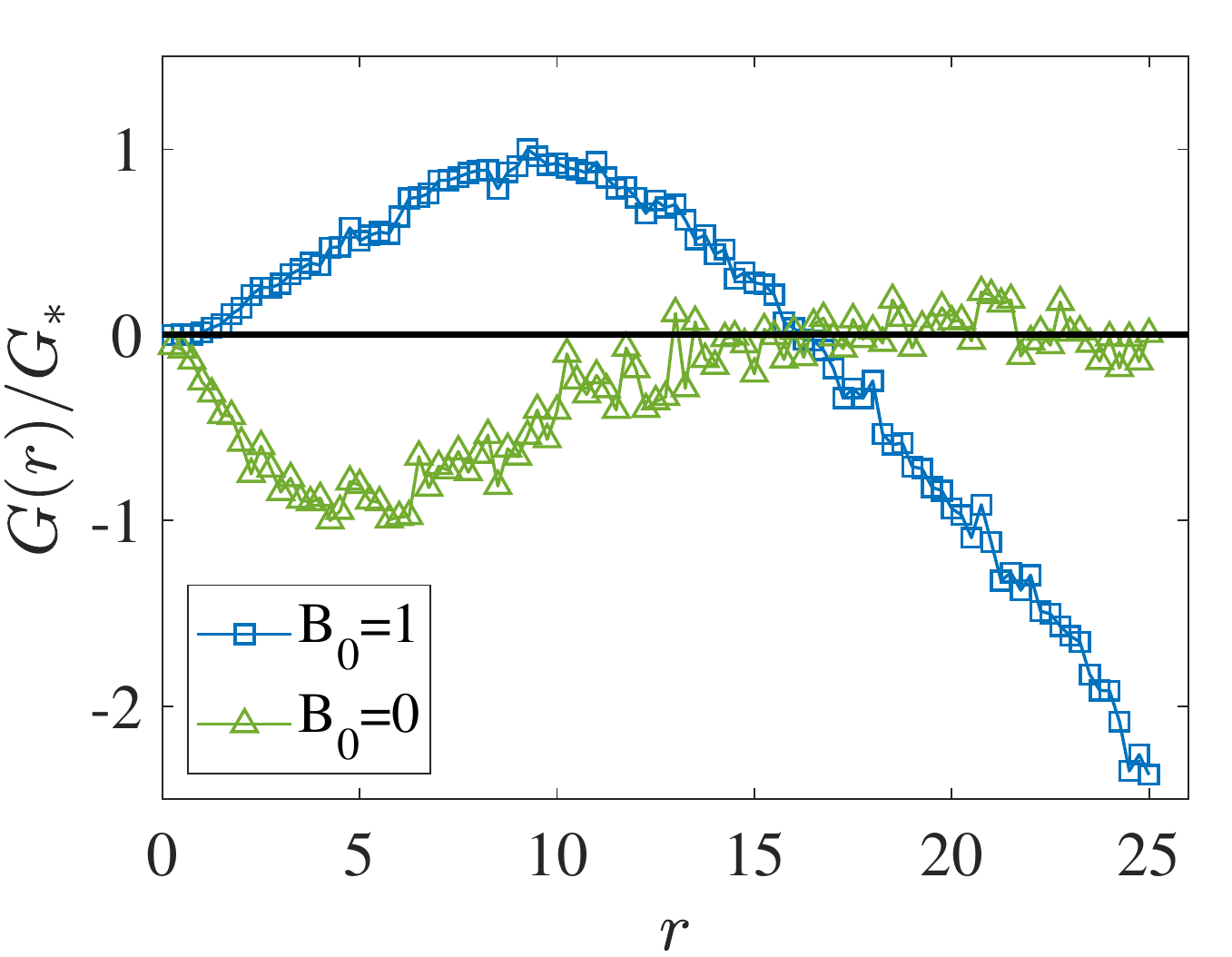}
\put(-200,137){(a)}\\
\includegraphics[trim=3.2cm 9.5cm 4cm 9.5cm, clip=true, scale=0.5]{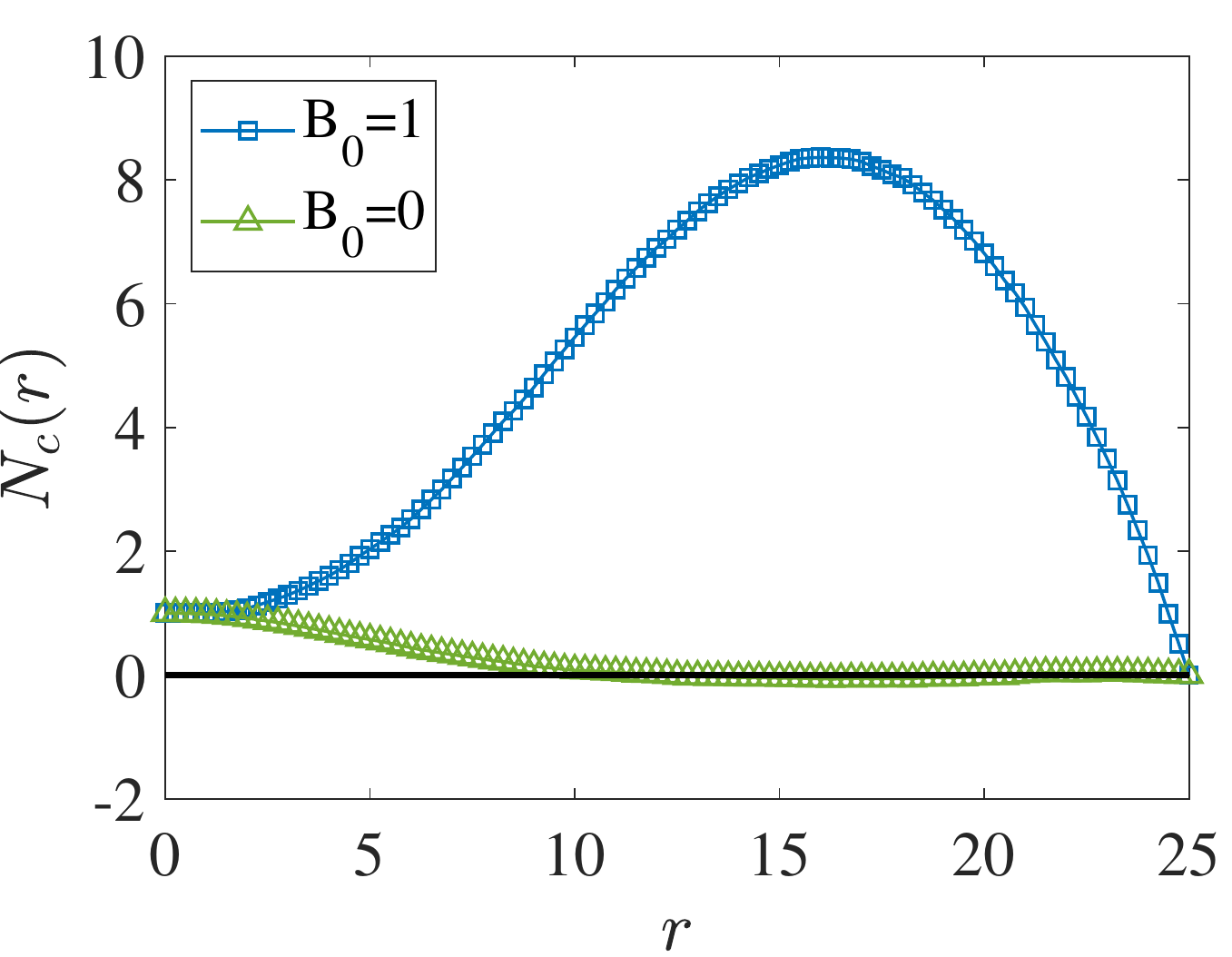}
\put(-200,140){(b)}
\caption{{Vortex correlation functions $G(r)/G_{*}$ and net vorticity $N_c(r)$ with and without magnetic fields in the initial state.} (a) Vortex correlation functions $G(r)/|G_{*}|$ for $B_{0}=0$ and $B_{0}=1$, where $G_*$ are scaling constants. From the different behaviors of $G(r)$, we can clearly identify the positive correlations between vortices in the FTM ($B_0=1$), while the vortices are negatively correlated in KZM ($B_0=0$);  (b) Net vorticity $N_c(r)$ inside a square for $B_{0}=0$ and $B_{0}=1$.  $N_c(r\lesssim16)>1$ indicates the positive correlations between vortices for $B_0=1$, while $N_c(r)<1$ indicates negative correlations for $B_0=0$.  Quench rate for both panels is $\tau_Q=20$ and we have made $200$ times of independent simulations for both panels. }   \label{vv}
\end{figure}

{\bf Spatial correlations \& net vorticity:} Spatial vortex correlation functions $G(r)$ and the net vorticity $N_c(r)$ can be used to quantitatively distinguish the correlation properties between vortices \cite{Hindmarsh:2000kd,golubchik2010,golubchik2011}. $G(r)$ is defined as $G(r)\equiv\langle n(r)n(0)\rangle$, with $n=+1(-1)$ at the location of a positive(negative) vortex, otherwise 0 elsewhere,
 and  $G(r)$ is calculated averagely over all vortices. In detail,  first, we can put one positive vortex at the origin, and then count the net vorticity at the various circumferences with distance $r$ to this origin; Then, we can choose another positive vortex as the origin, and to compute $G(r)$ with respect to it again; We do this procedure again and again by putting all the positive vortices as the origin. This is the procedure we did for one snapshot of the vortices in an independent run. We then choose another snapshot of the vortices in another independent run, and do the same procedure as above. And so on. We totally simulated 200 snapshots of the vortices. Finally, we average all the values of $G(r)$ at each distance $r$. In short,
 if there are $p$ positive vortices and $q$ negative vortices at the circumference, then $G(r)=p-q$. Obviously, if $G(r)$ is positive(negative) at short distance, it means the vortices are positively correlated(negatively correlated). In the panel (a) of Fig.\ref{vv} we exhibit the correlation functions $G(r)/G_*$ in the presence of magnetic field ($B_0=1$) and without magnetic field ($B_0=0$) for comparison. $G_*$ are constants to scale the amplitudes of $G(r)$ in order to compare the two cases easily. From the definition we can set $G(0)=0$. For $B_0=1$ we choose $G_*$ the maximum value of $G(r)$ while for $B_0=0$ we choose the absolute value of the minimum value of $G(r)$. We need to emphasize that a total scaling of $G(r)$ by dividing $G_*$ does not change the essence of correlations between vortices. For $B_0=0$ in the panel (a) in Fig.\ref{vv}, we clearly see that the vortices are negatively correlated in short range, which is a typical result from KZM. This is already well studied in previous work \cite{Zeng:2019yhi}. However, for $B_0=1$ we see that in the range $r\lesssim16$ the correlation function $G(r)$ is positive, indicating the vortices are positively correlated. This is a typically distinct character of the FTM from KZM. As $r$ is bigger, $G(r)$ becomes negative which can be easily understood from Fig.\ref{B} that at large $r$ there are more vortices with opposite sign.

 Another way to identify the positive correlations between vortices is to compute the net vorticity $N_c(r)$ {\it inside} the above square. \footnote{Be aware of the different definitions of $G(r)$ and $N_c(r)$. $G(r)$ is defined by counting the net vorticity at the circumference of the square, while $N_c(r)$ is defined within the square. } Panel (b) of Fig.\ref{vv} shows the different behaviors of $N_c(r)$ for the case of $B_{0}=0$ and  $B_{0}=1$, respectively. $B_0=0$ is for the KZM in which vortices are negatively correlated. Therefore, $N_c(r)$ decreases from $N_c(0)=1$ to zero at large distance. However, for $B_0=1$ the magnetic field plays an important role, the vortices are positively correlated from FTM. Therefore, $N_c(r)$ increases as $r$ becomes bigger, and then reach a maximum at around $r\approx16$. After the maximum it decreases to zero at very large distance. The behavior of $N_c(r)$ for $B_0=1$ is consistent with $G(r)$ in the panel (a), and it demonstrates that the vortices are positively correlated at a relatively long range which is a typical character of the FTM.


\begin{figure}[h]
\centering
\includegraphics[trim=3.2cm 9.5cm 4cm 9.5cm, clip=true, scale=0.5]{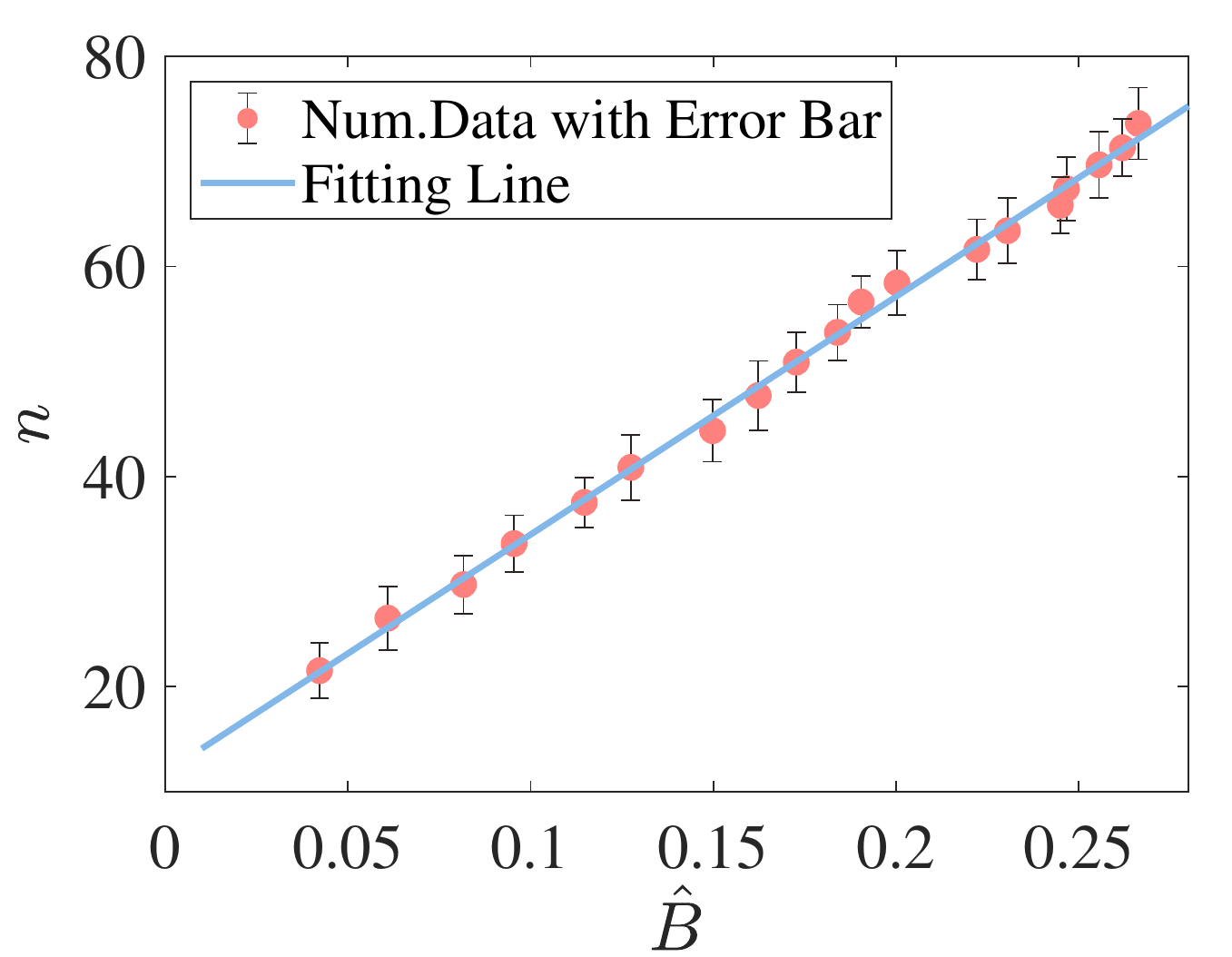}
\caption{Linear relation between the vortex number and the magnetic field at the `trapping' time.  Red dots are numerical data while blue lines are from the best fit. Error bars denote the standard deviations. We count the vortex number $n$ as the order parameter saturates equilibrium, and this relation is averaged over $200$ times of independent simulations. }\label{B0}
\end{figure}
{\bf Vortex number \& magnetic field:} From FTM, vortex number is proportional to the absolute value of magnetic fluxoids at the time that flux trapping takes place. Let's denote this time as `trapping' time $\hat t$, the flux at this moment as $\hat \Phi$ and magnetic field as $\hat B$. Thus, $n\propto|\hat\Phi|/\Phi_0$, in which $\Phi_0=2\pi$ is the fundamental magnetic fluxoids quantum. In our case the magnetic field has the plane-wave form in the initial state, then quenching induces the overdamping of the magnetic field with the amplitude decaying as $B(t)=e^{-\gamma t}$ (where $\gamma\approx k^2\approx 0.016$ \cite{Stephens:2001fv}) until the order parameter becomes relevant  (see \hyperref[app]{Appendix} for this exponential decay). From \hyperref[app]{Appendix} we recognize that during the overdamping the magnetic field maintains its plane-wave form, i.e. the wave number $k$ does not change, while only the amplitude decays. The `trapping' time $\hat t$ occurs at the instant that the amplitude departs away from this exponential decay.
Therefore, according to FTM, vortex number $n$ should be proportional to the amplitude of magnetic field at the `trapping' time, since $n\propto|\hat \Phi|=\int |\hat B\cos(kx)|dxdy\propto \hat B$. This linear relation is reflected in Fig.\ref{B0}, in which we quench the system with various initial amplitudes of magnetic field $B_0$ while fixing the quench rate as $\tau_Q=20$. This linear relation between $n$ and $\hat B$ provides a strong evidence to FTM.

\section{Conclusions}
By virtue of AdS/CFT correspondence, we achieved the clusters of strongly coupled equal-sign vortices from the FTM, which was a distinct mechanism compared to KZM in forming topological defects when local gauge symmetry was important.  Quenching the system into a superconductor phase, clusters of equal-sign fluxoids emerged from the initial plane-wave magnetic fields.
Vortex correlation functions and the net vorticity inside a loop quantitatively supported our findings. Linear dependence of the vortex number to the magnetic field at the `trapping' time demonstrated the FTM very well.  Although our model was in two dimensional space, which was not a good approximation for a superconductor film, it would provide a tractable and interesting model to examine the importance of magnetic fluctuations on critical dynamics and defect formations, which may be easily performed in superconductor experiments.

\section*{Acknowledgements}
This work was partially supported by the National Natural Science Foundation of China (Grants No. 11675140, 11705005 and 11875095) and supported by the Academic Excellence Foundation of BUAA for PhD Students.

\widetext

\section*{---Appendix---}
\label{app}


\setcounter{equation}{0}
\setcounter{figure}{0}
\setcounter{table}{0}
\setcounter{section}{0}
\makeatletter
\renewcommand{\theequation}{S\arabic{equation}}
\renewcommand{\thefigure}{S\arabic{figure}}
\renewcommand{\bibnumfmt}[1]{[#1]}
\renewcommand{\citenumfont}[1]{#1}

\section{Exponential decay of magnetic field in the early stage}
\label{Bdecay}

\begin{figure*}[htbp]
\centering
\includegraphics[trim=2.7cm 7.cm 4.cm 8.cm, clip=true, scale=0.36]{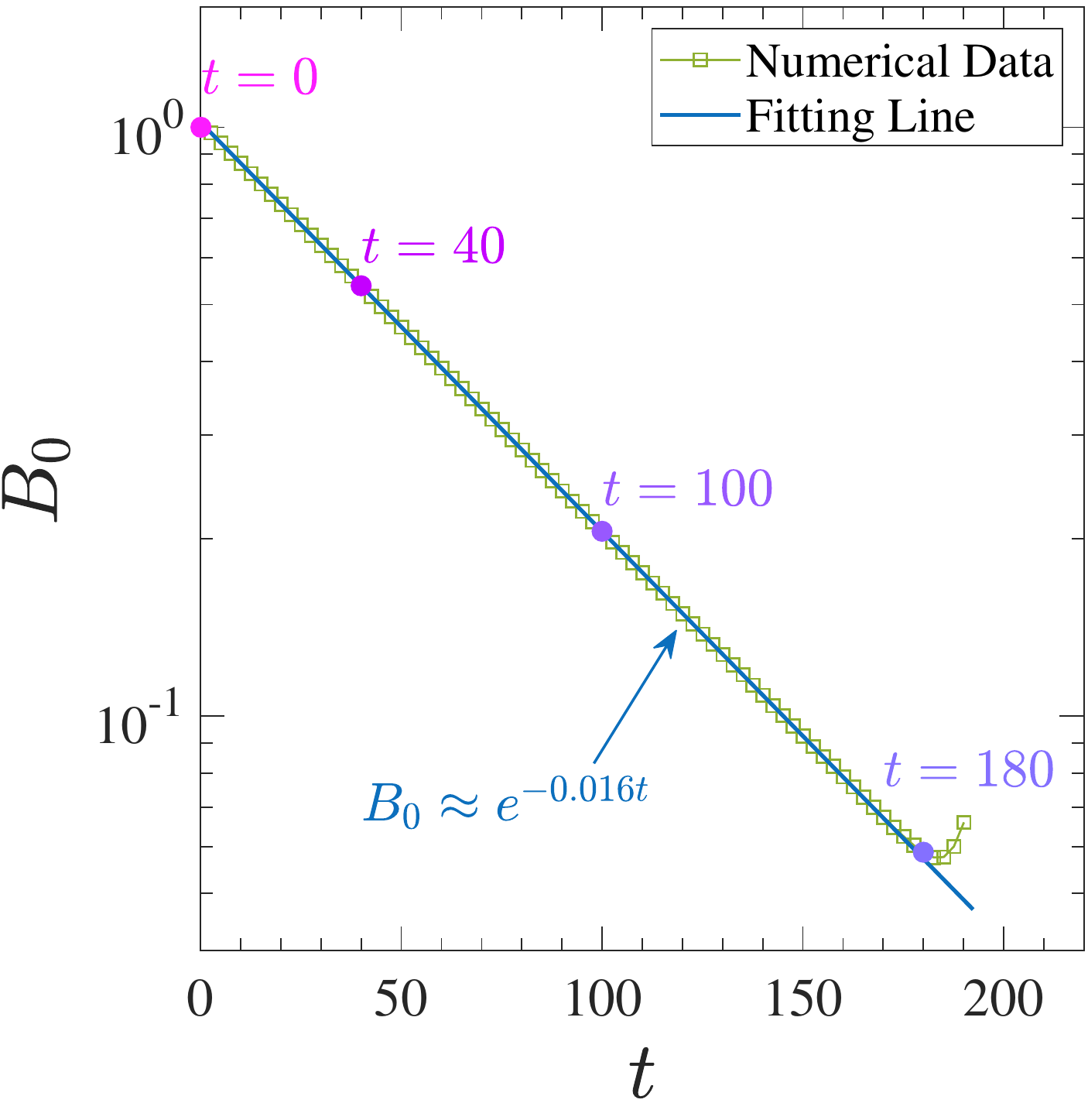}~~
\includegraphics[trim=1.5cm 0.5cm 0.8cm 0.8cm, clip=true, scale=0.38]{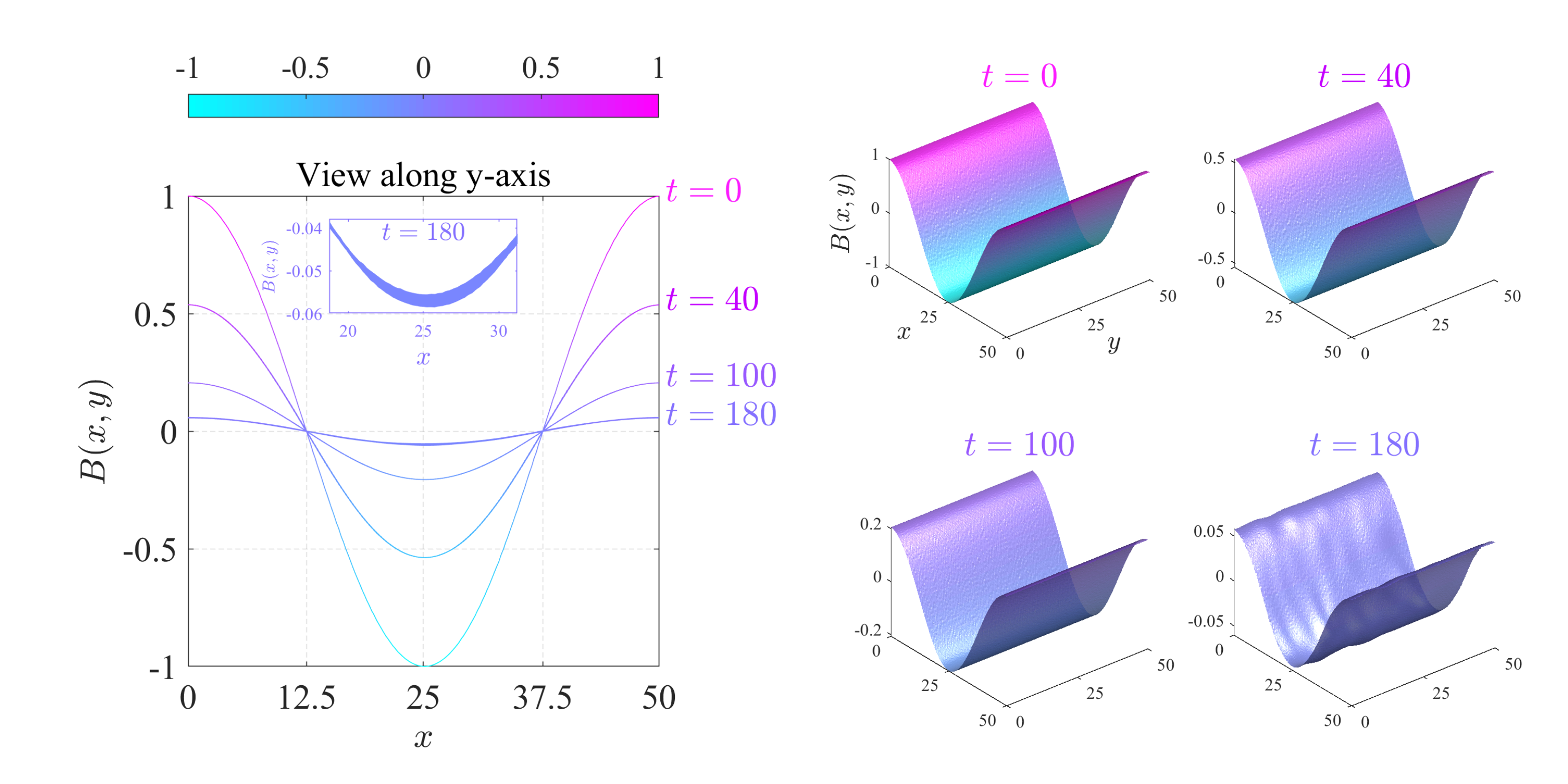}
\put(-470,150){(a)}
\put(-315,150){(b)}
\put(-155,150){(c)}
\caption{The exponential decay of magnetic field in the early stage with quench rate $\tau_Q=1000$ and initial amplitude of magnetic field $B_0(t=0)=1$. (a) Logarithmic plot of the time evolution of the amplitude of magnetic field. The linear dependence indicates an exponential decay of amplitude as $B_0(t)=B_0(t=0)e^{-\gamma t}$.  The four instants $(t=0, t=40, t=100,$ $  t=180)$ correspond to the four snapshots in the subsequent panels (b) and (c). Time $t=180$ is the `trapping' time that the amplitude deviates away from the exponential decay; (b) Configurations of the magnetic field at the four instants from the view parallel to $y$-axis. At the instants $(t=0, t=40, t=100)$ the magnetic fields still maintain in a perfect plane-wave form with wave number $k=2\pi/l$ where $l=50$. However, at the instant $t=180$ the shape of the magnetic field starts to deviate from the plane-wave from, which can be seen from the inset picture that near the bottom of the curve $(x=l/2)$ it becomes thicker. This thickness comes from the ripples that the magnetic field will start to form lumps and finally to shape the vortices; (c) 3D visualizations of the magnetic field at the four instants. At $(t=0, t=40, t=100)$ the magnetic fields are perfectly in the plane-wave shapes along $x$-axis and very smooth along the $y$-axis. However, at $t=180$ ripples turn out in the magnetic field since the order parameter gets bigger and ruins the exponential decay of the magnetic field. These ripples will finally form lumps and then vortices of the magnetic field, which is the core of FTM.}\label{Btime}
\end{figure*}

In Fig.\ref{Btime} we show the exponential decay of the amplitude of magnetic field in the early stage and the onset of the flux trapping. Fig.\ref{Btime} shows an example of $B_0(t=0)=1$ and $\tau_Q=1000$. Other values of parameters are similar as we have checked, which means the exponential decay in the early stage are independent of quench rate and the initial amplitude of the magnetic field. The decay rate in $B_0(t)=B_0(t=0)e^{-\gamma t}$ is always $\gamma\approx k^2\approx 0.016$ as stated in \cite{Stephens:2001fv}. 

The four instants $(t=0, t=40, t=100, t=180)$ as denoted in panel (a) are corresponding the four snapshots in the subsequent panels (b) and (c). From panels (b) and (c) we find that at the instants $(t=0, t=40, t=100)$ the magnetic fields are still in plane-wave form (along $x$-direction) perfectly. The only difference is that their amplitudes decrease according to $B_0(t)=B_0(t=0)e^{-\gamma t}$.

However, at instant $t=180$ the amplitude of magnetic field will deviate away from the initial exponential decay (see panel (a)), since in this case the effect of the scalar field cannot be ignored. The effect of scalar field is complicated which could be only studied by numerics as we already showed in the main text.  From panels (b) and (c) we indeed see that the magnetic field will start to be away from plane-wave form. Some ripples appear in the magnetic field at this instant (panel (c)). This is the onset of the flux trapping. Thus, $t=180$ is the `trapping' time as we called. These ripples will finally become lumps in the magnetic field, and then turn out to be vortices when system goes to the equilibrium state, as we already showed in the Fig.2 in the main text.

\section{Locations of the magnetic fields and the order parameter vortices in the far-from-equilibrium state}
\label{location}
\begin{figure*}[h]
\centering
\includegraphics[trim=3.5cm 9.cm 3.8cm 9cm, clip=true, scale=0.5]{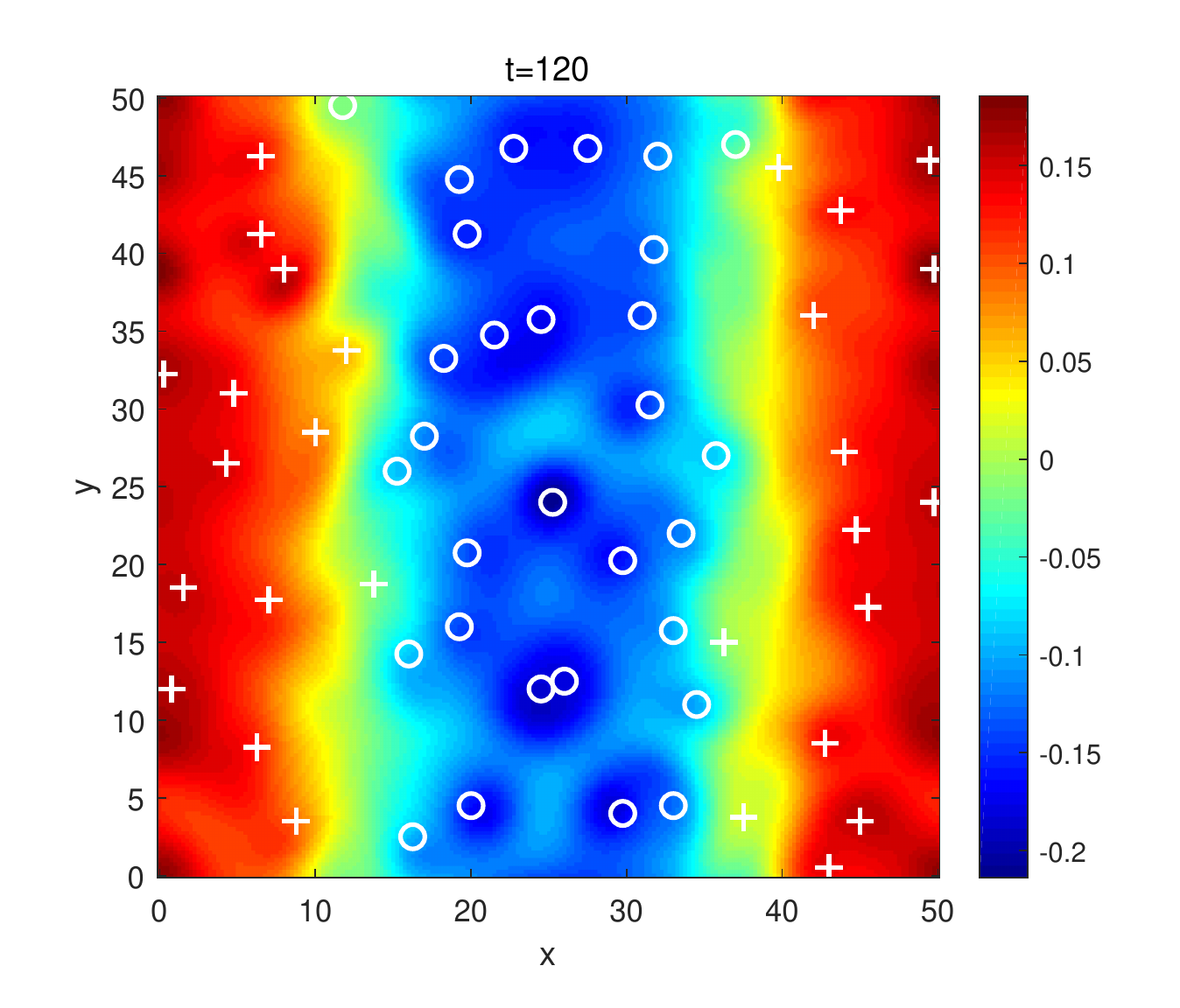}
\includegraphics[trim=3.2cm 9.cm 4cm 9cm, clip=true, scale=0.5]{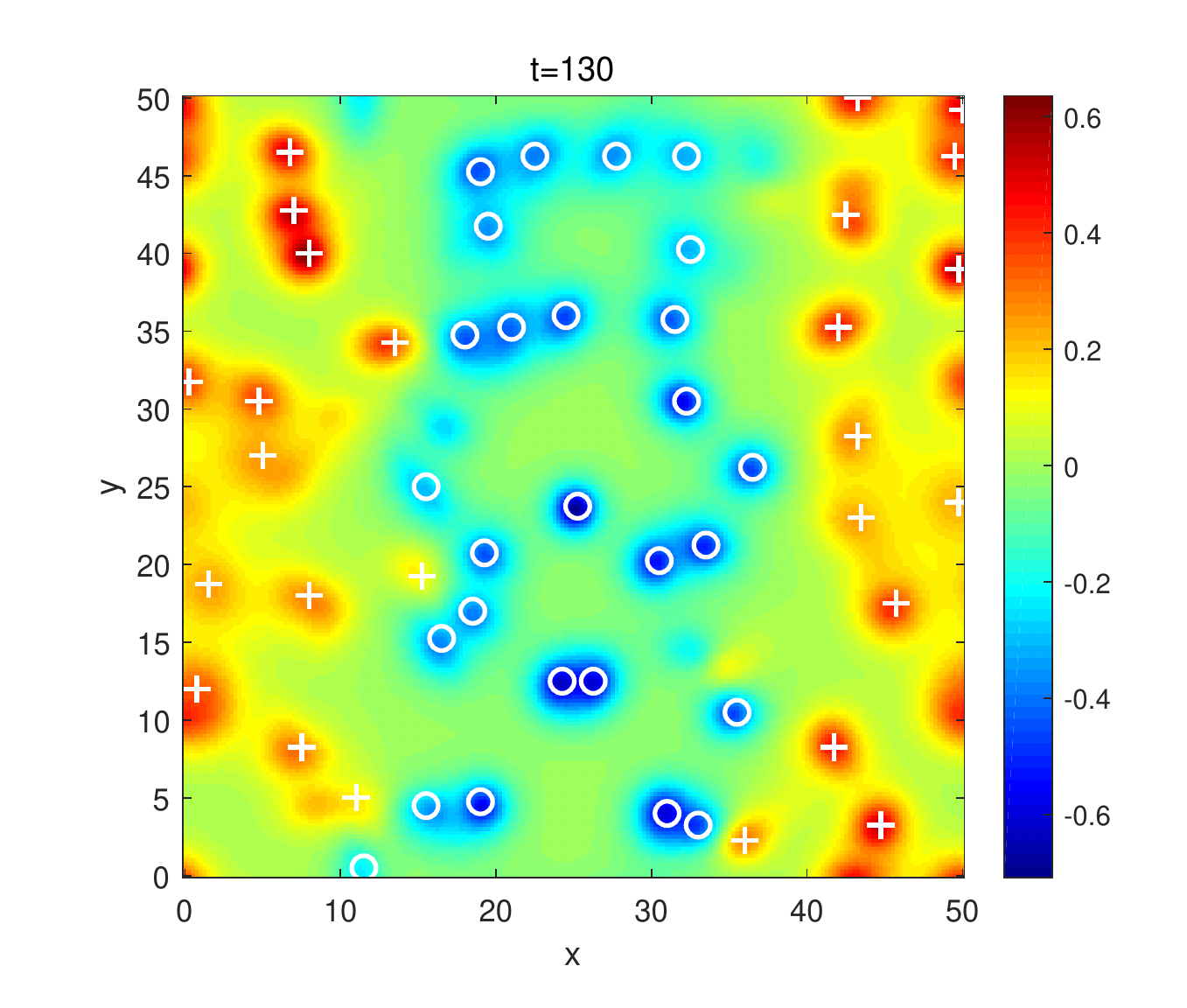}
\caption{ Density plots of the magnetic fields and the locations of the singular points of the order parameter phases (i.e. the locations of the centers of the vortices) at two specific times $t=120$ (left) and $t=130$ (right) in the far-from-equilibrium state with $\tau_Q=20$ and $B_0=1$. Colorful regions represent the magnitudes of the magnetic fields, while the plus signs and the circles indicate the positions of the positive and negative vortices, respectively.}\label{BO}
\end{figure*}

Theoretically, from the knowledge of type-II superconductor,  there is a gauge invariant term, such as $\nabla\varphi-\bf{A}$, where $\varphi$ is the phase of the order parameter while $\bf{A}$ is the spatial components of the gauge field, in the free energy.  Therefore, the magnetic field $B=\nabla\times\bf{A}$ will make frustrations in the phase of order parameter. This leads to the phenomenon that the locations of the singular points of the order parameter phases will correspond to the finite values of magnetic fields. This is why the location of magnetic fluxoids and the order parameter vortices are at the same places.

In the far-from-equilibrium state, the fluxes of magnetic fields are not quantized. Therefore, we can only see the `condensate' or `lumps' of the magnetic fields  in the far-from-equilibrium state. In Fig.\ref{BO}, we numerically show the density plots of the magnetic fields and the locations of the singular points of the order parameter phases (i.e. the locations of the centers of the vortices) for time $t=120$ and $t=130$. Times $t=120$ and $t=130$ are in the far-from-equilibrium state, which can be found in the Fig.\ref{B} or from the movie \href{https://bhpan.buaa.edu.cn:443/link/95BA4D4BD5B130767C9012BFAC4C91DB}{M1.avi}. From the Fig.\ref{BO}, we can find most of the order parameter vortices are located at places of the condensates or lumps of the magnetic fields. Some minor vortices are not sitting at the places of the condensates of the magnetic fields at time $t=120$. But it is understandable that they are in the far-from-equilibrium state, as times goes by, these minor vortices will soon disappear, for instance, at time $t=130$.

\newpage


\end{document}